\documentclass[twocolumn,aps,a4]{revtex4}
\usepackage{graphicx}
\usepackage{dcolumn}
\usepackage{amsmath}

\begin{document}

\title{The pressure dependence of electron-phonon coupling in
the organic superconductor $\kappa$-(BEDT-TTF)$_{2}$Cu(SCN)$_{2}$: A
comparison of high pressure infrared reflectivity and Raman
scattering experiments.}

\author{R.~D.~McDonald, A-K.~Klehe, J.~Singleton and W.~Hayes}
\address{Clarendon Laboratory, Department of Physics, Parks Road, Oxford, OX1~3PU, UK.}

\date{May 2001}

\begin{abstract}
We determine the pressure dependence of the electron-phonon coupling
constant in $\kappa$-(BEDT-TTF)$_{2}$Cu(SCN)$_{2}$ by comparison of
high pressure Raman scattering and high pressure infrared
reflectivity (IR) measurements. We use comparison of the IR
reflection spectrum from deuterated and protonated samples to aid
the deconvolution of several overlapping phonon modes. These coupled
modes are modelled with a Green's function to extract the linear
pressure dependence of the individual modes. The Raman active
molecular vibrations of the BEDT-TTF dimers stiffen by
0.1-1~$\%$GPa$^{-1}$. In contrast, the corresponding modes in the IR
spectrum are observed at lower frequency, with a pressure dependence
of 0.5-5.5~$\%$GPa$^{-1}$, due to the influence of the
electron-phonon interaction. The stronger pressure dependence of the
central C=C mode of the BEDT-TTF molecule in the IR, is discussed.
Our analysis suggests that reduction of electron-phonon coupling
under pressure does not account for observed suppression of
superconductivity under pressure.
\end{abstract}

\maketitle
\newpage
\begin{sloppypar}

$\kappa$-(BEDT-TTF)$_{2}$Cu(SCN)$_{2}$ is one of the best
characterized organic superconductors~\cite{Singleton1}. It is a
highly anisotropic material with a quasi-two dimensional band
structure, whose Fermi-surface topology has been determined by
magnetotransport experiments~\cite{Singleton1,Jason}. At ambient
pressure $\kappa$-(BEDT-TTF)$_{2}$Cu(SCN)$_{2}$ is a superconductor
with a transition temperature of $T_{\rm c}\simeq 10.4$~K. $T_{\rm
c}$ decreases upon the application of pressure until, at pressures
$P$ exceeding 0.5~GPa, superconductivity is fully
suppressed~\cite{Jason,PT}. The effective mass, $m^{*}$, derived
from magnetic quantum oscillation measurements, decreases linearly
with pressure up to 0.5~GPa; above this pressure the magnitude of
d$m^{*}$/d$P$ is strongly reduced~\cite{Jason}. In contrast, the
effective mass derived from optical measurements, $m_{opt}$,
decreases approximately linearly throughout this pressure
range~\cite{pressIR}. The coincidence of a ``kink'' in the pressure
dependence of $m^{*}$ with the pressure above which
superconductivity is suppressed and the absence of a ``kink''in the
pressure dependence of $m_{opt}$, indicates that the interactions
parameterised by $m^{*}$ are connected to the superconductivity.

The key question to address is the effect of these interactions,
{\it i.e.} what is the dominant pairing mechanism for
superconductivity in this material? In this paper we compare
infrared (IR) \cite{pressIR} and Raman scattering \cite{pressRaman}
measurements under pressure to determine the role of the
electron-phonon interaction. This is possible because the IR
measurement probes the molecular vibrations dressed by the
electron-phonon interaction \cite{Kornelsen1,Sugano1}, whereas
non-resonant Raman measurements probe the bare mode frequencies
\cite{Sugano1,Rice1}. Our recent high-pressure IR
study~\cite{pressIR} of this material revealed anomalously large
mode stiffening for the molecular vibrations most strongly coupled
to the electronic excitations. However, determination of the
pressure dependence of the most strongly coupled mode, at around
1300~cm$^{-1}$ in the IR spectrum, is not straightforward due to the
mixing of several modes with varying pressure dependence. In the
current paper we use the comparison of ambient pressure reflectance
data from protonated and deuterated samples to establish a coupled
oscillator model. This model is then employed to determine pressure
induced frequency shifts of the interacting modes. By employing a
dimer charge-oscillation model \cite{Sugano1,Rice1,aprox},
comparison of the pressure induced frequency shifts observed in both
the IR \cite{pressIR} and the Raman \cite{pressRaman} spectra
determines the pressure dependence of the electron-phonon
interaction. Four modes are observable in both the high-pressure
Raman and IR spectra (see Table \ref{wtable}). They are labeled with
subscripts indicating the atoms/bond predominantly involved in the
vibration \cite{EldridgeET}. In order of increasing frequency they
are: the C-S mode originating from the BEDT-TTF 60~B3g asymmetric
vibration, the central C=C mode originating from the BEDT-TTF
3~A$_{g}$ symmetric vibration and two Cu(SCN)$_{2}$ anion modes.

\begin{figure}[ht]
\includegraphics[width=9cm]{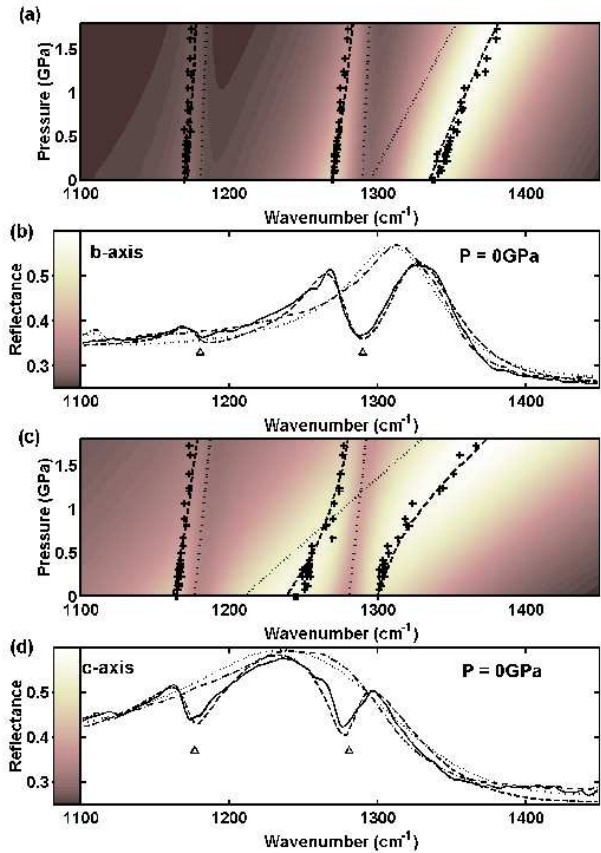}
\centering  \caption{ {\scriptsize (a) and (c) diamond/sample
reflection surfaces generated from the coupled oscillator model for
the {\it b} and {\it c} axis respectively; the reflectance scales
are adjacent to the vertical axes of (b) and (d). Superimposed are
the frequency of reflectance peaks (crosses) \cite{pressIR},
uncoupled mode frequencies (dotted lines) and the peaks of the
coupled oscillator model (dashed lines). (b) and (d) The ambient
pressure reflection spectrum for polarization parallel to the {\it
b} and {\it c} axes respectively: solid and dashed lines for the
protonated salt (measurement and fit), dash-dot and dotted lines for
the deuterated salt (measurement and fit).}} \label{FIG1}
\end{figure}

Owing to its strong electron-phonon coupling \cite{Sugano1}, we are
particularly interested in extracting the pressure induced frequency
shift of the central C=C mode. Dimerization in the $\kappa$-phase
makes this mode IR active \cite{Sugano1,EldridgeIR1}.
Electron-phonon coupling softens this mode from its bare, Raman
active, frequency of 1470~cm$^{-1}$ to around 1290~cm$^{-1}$ along
the {\it b}-axis and around 1220~cm$^{-1}$ along the {\it c}-axis.
This brings the mode into resonance with the end carbon-hydrogen
(C-H) oscillations of the BEDT-TTF molecule \cite{IRCH1}. Upon
deuteration the C-H(D) modes are softened sufficiently to reveal the
true line shape of the central C=C mode. Comparison of the IR
spectrum from the protonated and deuterated salts illustrates the
structure of the C=C mode due to this coupling (see Fig.~\ref{FIG1}b
and \ref{FIG1}d); dips occur where the stronger C=C mode loses
spectral weight to the weaker C-H modes it is driving.

Although mode mixing should be treated from a quantum mechanical
point of view, it has been shown \cite{fermires1} that a classical
approach, such as described below, can yield information regarding
the individual mode frequencies. From the equations of motion
(Fourier components) for a set of $N$ coupled oscillators
\cite{Barker}, a Green's-function matrix approach \cite{Ryan} is
employed to model the dielectric response function
\begin{equation}
\epsilon(\omega) = \Omega_p \{G(\omega)\}^{-1} \Omega_p^T,
\label{green1}
\end{equation}
where $\Omega_p$ is a vector containing each oscillator's spectral
weight, $\Omega_p^T$ is its transpose and $G$ is a square matrix of
dimension $N$, with $N$ the number of oscillators.  The diagonal
terms in $G$ are of the form
\begin{equation}
G_{aa}(\omega) = (\omega_a^2 + i\omega\Gamma_a - \omega^2),
\end{equation}
specifying each Lorentzian oscillator of frequency $\omega_{a}$ and
damping $\Gamma_{a}$. The off-diagonal terms in $G$ provide the
coupling and are of the form
\begin{equation}
G_{ab}(\omega) = (\Delta_{ab}^2 + i\omega\Gamma_{ab}). \label{3}
\end{equation}
This coupled oscillator model is employed to analyze our room
temperature high-pressure IR reflectance \cite{pressIR}. The aim is
to extract the pressure-induced frequency shift of individual modes
from the apparently non-linear shift (predominantly in the {\it
c}-axis response) of the reflectance peaks as a function of pressure
(the crosses in Fig.~\ref{FIG1}a and Fig.~\ref{FIG1}c).

To restrict the possible parameter space encountered, a simple
(uncoupled) Drude-Lorentz oscillator fit to the ambient-pressure
room-temperature IR spectrum of the deuterated salt was used as a
starting point, (the dotted line in figure~\ref{FIG1}b and
\ref{FIG1}d). In the absence of coupling the asymmetric line shape
of the central C=C mode for both {\it b} and {\it c} axes is most
accurately reproduced when the Drude term is heavily suppressed,
{\it i.e.} damped to the point that it hardly contributes to the
reflectance in this frequency range. The terms describing the
background and C=C mode were then held fixed and free parameters
describing the two C-H modes and their coupling terms were
introduced. A least squares fitting procedure was employed, yielding
the parameters necessary to model the interacting modes in the
protonated salt at room temperature and ambient pressure (the dashed
and solid line in figures~\ref{FIG1}b and \ref{FIG1}d respectively).

For both {\it b}- and {\it c}-axis polarizations and both the
protonated and deuterated samples the damping and strength of the
two C-H(D) modes tended to the same values, $\Gamma_{H(D)}$
$\rightarrow$ 16~cm$^{-1}$ and $\Omega_{p,H(D)}$ $\rightarrow$ 0,
{\it i.e.} two sharp modes with no infrared activity of their own.
The coupling strengths, $\Delta$ (see equ.(\ref{3})), are listed in
table~\ref{Dtable}, with the cross damping term $\Gamma_{ab}$ also
tending to zero for all modes.
\begin{table}[ht]
\begin{center}
\begin{tabular}{ccc} \hline \hline
Parameter & b-axis value & c-axis value\\
 & cm$^{-1}$ & cm$^{-1}$\\
  \hline\hline
   $\Delta_{\rm CC,CH_{1}}$ & 260 & 230\\
    $\Delta_{\rm CC,CH_{2}}$ & 260 & 230\\
     $\Delta_{\rm CH_{1},CH_{2}}$ & 160 & 0\\
      \hline
\end{tabular}
\caption{mode coupling parameters.} \label{Dtable}
\end{center}
\end{table}
With all parameters except the three mode frequencies held constant,
the pressure-induced frequency shifts of the reflectance peaks were
fitted.  The data could be successfully modelled with only first
order (linear) pressure shifts included (see Fig.~\ref{FIG1}a and
\ref{FIG1}c and table~\ref{wtable}).
\begin{table}[h]
\begin{center}
\begin{tabular}{cccc} \hline \hline
 Mode & Raman & Infrared b-axis & Infrared c-axis\\
 & cm$^{-1}$+(\%GPa$^{-1}$) & cm$^{-1}$+(\%GPa$^{-1}$) & cm$^{-1}$+(\%GPa$^{-1}$)\\
  \hline\hline
  $\omega_{\rm CT}$ & ------- & 2910 + 4.0 & 2390 + 4.0\\
   $\omega_{\rm CS}$ & 886.2 + 0.85& 883.5 + 0.71 & 873.6 + 1.0\\
   $\omega_{\rm CC}$ & 1467.7 + 0.4& 1290 + 2.5 & 1210 + 5.5\\
    $\omega_{\rm CH_{1}}$ & ------- & 1181 + 0.5 & 1177 + 0.5\\
     $\omega_{\rm CH_{2}}$ & ------- & 1290 + 0.5 & 1281 + 0.5\\
     $\omega_{\rm CD_{1}}$ & ------- & 1027 $^{\dag}$  & 1027 $^{\dag}$ \\
     $\omega_{\rm CD_{2}}$ & ------- & 1116 $^{\dag}$  & 1122 $^{\dag}$ \\
     $\omega_{\rm anion1}$ & 2064.6 + 0.1 & 2067.4 + 0.1 & 2065.6 + 0.15\\
     $\omega_{\rm anion2}$ & 2106.3 + 0.2 & ------- & 2109.3 + 0.2\\
      \hline
\end{tabular}
{\scriptsize $^{\dag}$ No infrared pressure data are available for
the deuterated salt.}\\ \caption{Raman and IR frequencies and
pressure shifts \cite{pressIR,pressRaman}.} \label{wtable}
\end{center}
\end{table}

Including a pressure-dependent Drude response to account for the
increase in background reflectance in this spectral region has a
small effect ($\approx$~+0.5~\%) on the observed peak frequencies in
the reflectance model. Its main effect is to broaden and distort the
line shape of the phonon modes towards higher frequency, as
experimentally observed at high pressure \cite{pressIR}.

Within the framework of a dimer charge oscillation model
\cite{Rice1} it is possible to calculate a dimensionless
electron-phonon coupling constant, $\lambda$, using a mode's IR
frequency, $\omega_{\rm IR}$, and the corresponding Raman frequency,
$\omega_{\rm R}$, given knowledge (see Table~\ref{wtable}) of the
energy, $\omega_{\rm CT}$, of the coupling charge transfer band
\cite{Sugano1,aprox}
\begin{equation}
\frac{\omega_{\rm R}^{2}-\omega_{\rm IR}^{2}}{\omega_{\rm R}^{2}}
\simeq \lambda\frac{\omega_{\rm CT}^{2}}{\omega_{\rm
CT}^{2}-\omega_{\rm R}^{2}} \label{lameq}
\end{equation}
(the C=C mode, being an asymmetric combination of A$_{g}$ molecular
modes, couples to the intra-dimer charge transfer \cite{Kozlov}).
The dominant source of error in this calculation is due to the width
of $\omega_{\rm CT}$. The pressure derivative of equation
(\ref{lameq}),
\[
\frac{{\rm d}\ln\lambda}{{\rm d}P} = 2\frac{\omega_{\rm
IR}^{2}}{\omega_{\rm R}^{2}-\omega_{\rm IR}^{2}}\left[\frac{{\rm
d}\ln\omega_{\rm R}}{{\rm d}P}-\frac{{\rm d}\ln\omega_{\rm IR}}{{\rm
d}P}\right]
\]
\begin{equation}
+ 2\frac{\omega_{\rm R}^{2}}{\omega_{\rm CT}^{2}-\omega_{\rm
R}^{2}}\left[\frac{{\rm d}\ln\omega_{\rm CT}}{{\rm d}P}-\frac{{\rm
d}\ln\omega_{\rm R}}{{\rm d}P}\right], \label{lamdir}
\end{equation}
has a weaker dependence on the value of $\omega_{\rm CT}$. However
we found that the dominant error in (\ref{lamdir}) arises from the
difficulty in determining an accurate pressure dependence of the
broad CT band. Constraining the pressure dependence of $\omega_{\rm
CT}$ to 4$\pm$4~\%GPa$^{-1}$ for both {\it b}- and {\it c}-axes
\cite{pressIR}, results in an error in the pressure derivative of
the coupling constant for the C=C mode of $\pm$3.5~\%GPa$^{-1}$ for
the {\it b}-axis and $\pm$5~\%GPa$^{-1}$ for the {\it c}-axis. The
error in the pressure derivative of the coupling constant for the
C-S mode is larger (8~\%GPa$^{-1}$ for the {\it b}-axis and
28~\%GPa$^{-1}$ for the {\it c}-axis) because $\lambda_{\rm CS}$ is
so small. The Cu(SCN)$_{2}$ anion modes do not exhibit any evidence
of electron-phonon coupling, {\it i.e.} the difference between the
infrared and Raman frequencies for the anion modes is accounted for
by purely vibrational coupling. The zone center frequency
separation, $\Delta\omega_{\rm s}$, between the symmetric (Raman
active) and asymmetric (IR active) combinations of the molecular
vibrations, $\omega_{\rm mol}$, is determined by the frequency of
the optical branch of the lattice modes, $\omega_{\rm lattice}$
\cite{pressRaman,Zallen},
\begin{equation}
\Delta\omega_{\rm s}~=~\frac{\omega_{\rm lattice}^{2}}{2~\omega_{\rm
mol}}.
\end{equation}
The larger observed $\Delta\omega_{\rm s}$ for the anion mode in the
{\it b}-axis response (Table~\ref{wtable}), for which the lattice
mode is stiffer, is a further confirmation of the lattice mode
assignment given in~\cite{pressRaman}.

\begin{table}[h]
\begin{center}
\begin{tabular}{ccccc} \hline \hline
Mode & $\lambda_{b}$ & $\frac{{\rm d}\ln\lambda_{b}}{{\rm d}P}$ &
$\lambda_{c}$ & $\frac{{\rm d}\ln\lambda_{c}}{{\rm d}P}$\\
 & & \%GPa$^{-1}$ & & \%GPa$^{-1}$\\
  \hline\hline
   $\omega_{\rm CS}$ & 0.01(1) & 47.9 & 0.02(1) & -11.1\\
    $\omega_{\rm CC}$ & 0.17(1) & -11.9 &  0.20(1) & -17.3\\
     $\omega_{\rm anion1}$ & 0 & 0 & 0 & 0\\
     $\omega_{\rm anion2}$ & ------- & ------- & 0 & 0\\
      \hline
\end{tabular}
\label{Ltable} \caption{dimensionless electron-phonon coupling
constants and their pressure derivatives.}
\end{center}
\end{table}

Kozlov {\it et al.} \cite{Kozlov} calculate that the antiphase
combination of B$_{3g}$ molecular modes couple to charge transfer
perpendicular to the intra-dimer direction. This suggests that the
C-S mode couples to intra- not inter-band electronic transitions. It
should be noted however, that the electron-phonon coupling constants
for the C-S mode have been calculated using the same $\omega_{\rm
CT}$ as the C=C mode because it is impossible to distinguish the
contributions to the IR spectrum from inter- and intra-band
transitions.

It has been shown previously \cite{Yamaji} that the usual
electron-acoustic-phonon interaction mechanism is unable to account
for the magnitude of the electron-phonon coupling constant or the
large pressure dependence of $T_{\rm c}$ in the BEDT-TTF
superconductors. A further refinement \cite{Yamaji} is to include
the attractive interaction mediated by the A$_{g}$ molecular modes,
with the total electron phonon coupling constant, $\lambda_{\rm
TOT}$, given by a Yamaji sum over the individual A$_{g}$ molecular
modes. The energy scale for the interaction is still set by the
Debye frequency, $\Theta$ \cite{Yamaji,Jose1}. Caulfield {\it et
al.} \cite{Jason} have previously determined
$\Theta~\approx$~40cm$^{-1}$ by fitting the effective mass
dependence of $T_{\rm c}$ with a linearised Eliashberg equation
using an Einstein density of phonon states, $\delta(\Theta)$. High
and low temperature specific heat measurements \cite{C1,C2} yield
values of $\Theta$ ranging from 38cm$^{-1}$ to 140cm$^{-1}$.

Calculations of $\lambda_{\rm TOT}$ \cite{Yamaji} give values
ranging from 0.3-0.45 \cite{lambdaref1,Shumway}. Based on the
vibrations we sampled, we assume $\lambda_{\rm TOT}$ to have a
pressure dependence similar to that of the strongly coupled C=C
mode, {\it i.e.} of the order of -17~$\%$GPa$^{-1}$, with an upper
limit of -20$\%$GPa$^{-1}$ used in this calculation. The
weak-coupling BCS formula \cite{BCS1} gives a satisfactory
explanation of the ambient pressure $T_{\rm c}$ \cite{Sugano1} and
accurately describes the effective mass dependence of the
superconducting transition temperature \cite{Jason,disclamer}.

The pressure derivative of the weak-coupling BCS formula provides a
convenient parameterization of $\frac{{\rm d}\ln~T_{\rm c}}{{\rm
d}P}$ in terms of $\frac{{\rm d}\lambda}{{\rm d}P}$,
\begin{equation}
\frac{{\rm d}\ln~T_{\rm c}}{{\rm d}P} = \frac{{\rm d}\ln\Theta}{{\rm
d}P} + \frac{1}{(\lambda-\mu^{*})^{2}}\left[\frac{{\rm
d}\lambda}{{\rm d}P}-\frac{{\rm d}\mu^{*}}{{\rm d}P}\right].
\label{pressTc}
\end{equation}
With $\frac{1}{(\lambda-\mu^{*})^{2}} = [\ln(\frac{T_{\rm
c}}{1.13\Theta})]^{2}$, $\Theta~\approx$~90$\pm$50~cm$^{-1}$
\cite{Jason,C1,C2} and using the average pressure induced stiffening
of the Raman active lattice modes $\approx$~+13~\%GPa$^{-1}$
\cite{pressRaman} for $\frac{{\rm d}\ln\Theta}{{\rm d}P}$, the only
unknown is the pressure dependence of the Coulomb pseudopotential,
$\frac{{\rm d}\mu^{*}}{{\rm d}P}$. Calculations of $\mu^{*}$
\cite{Jason,Shumway} indicate a small positive value, a possible
indication that direct interactions between the quasiparticles are
involved in the pairing. There is also sufficient uncertainty in the
parameters necessary to calculate the pressure dependence of the
dimer-site Coulomb repulsion \cite{mckenzie} that no reliable
prediction can be made. We have thus decided to ignore $\frac{{\rm
d}\mu^{*}}{{\rm d}P}$ in the above calculation. This gives
$\frac{{\rm d}\ln~T_{\rm c}}{{\rm
d}P}~\approx$~-40$\pm32~\%$GPa$^{-1}$, which is far from the
experimentally observed value of $\frac{{\rm d}\ln~T_{\rm c}}{{\rm
d}P}~\approx$~-200~\%GPa$^{-1}$ \cite{Jason,PT}. As can be seen from
(\ref{pressTc}), $\Theta$ directly scales with the pressure
dependence of $T_{\rm c}$. To obtain the observed rapid fall of
$T_{\rm c}$ with pressure from only the decrease in the
electron-phonon coupling constant, requires $\Theta$ to be of the
order of the C=C mode frequency, $\approx$~1500~cm$^{-1}$. Such a
value is utterly inconsistent with the 10~K superconducting
temperature and the unconventional isotope shift observed upon
carbon substitution \cite{isotope}.

In conclusion we have shown that it is possible to recreate the line
shape of the resonance between the central C=C mode and the H (or D)
modes in protonated (or deuterated)
$\kappa$-(BEDT-TTF)$_{2}$Cu(SCN)$_{2}$. This was achieved by
coupling the strongly infrared active C=C mode to two C-H modes
which possessed negligible IR strength of their own. We have also
shown that the apparent non-linearity of the pressure dependence of
the modes is due to anti-crossing of the mixed modes. We compare
high-pressure Raman scattering and IR reflectivity data enabling the
pressure dependence of the electron-phonon coupling strength to be
evaluated for modes observed in both spectra. Using the weak
coupling limit of BCS theory we have been able to compare the
pressure dependence of the electron-phonon coupling constant and the
pressure dependence of the superconducting transition temperature.
This casts considerable doubt on whether this material is a simple
BCS superconductor because the characteristic energy of the pairing
interaction would have to be of the order of the highest frequency
molecular modes, a value inconsistent with the 10~K superconducting
transition temperature. This is an indication that electron-electron
interaction may be playing a significant role in this material's
superconducting mechanism.

The authors thank A.F.~Goncharov, V.V.~Struzhkin, R.J.~Hemley,
A.~P.~Jephcoat and H.~Olijnyk for their experimental support and
helpful discussions. This work was supported by the EPSRC.

\end{sloppypar}


\begin{thebibliography}{00}
\bibitem{disclamer}
The fact that the weak-coupling BCS expresssion describes $T_{\rm
c}$ versus $m^*$ well [2] should NOT be taken to imply that
$\kappa$-(BEDT-TTF)$_{2}$Cu(SCN)$_{2}$ is a weak-coupling BCS
superconductor. The formula is here used as a convenient
parameterisation which is known to describe earlier data well [2].
\end{thebibliography}
\end{document}